\documentclass[conference]{IEEEtran}
\IEEEoverridecommandlockouts
\usepackage{cite}
\usepackage{amsmath,amssymb,amsfonts}
\usepackage{algorithmic}
\usepackage{graphicx}
\usepackage{textcomp}
\usepackage{xcolor}
\usepackage{amsmath}% for double integral symbol in this template
\usepackage{graphicx}% for figures
\usepackage{multirow}% to allow multiple-row elements in tabular environment
\usepackage[none]{hyphenat}% turn off hyphenation to make text extraction and indexing easier
\usepackage{float}% better control of floating figures and tables
\usepackage{subfig}% for subfigures within figures
\usepackage{dblfloatfix}% fix to allow page-wide floats at bottom of page
\usepackage{t1enc}% allows access to various special characters
\usepackage{times}% change font to Nimbus Roman (based on Times Roman)
\usepackage[usestackEOL]{stackengine}
\usepackage[numbers,sort&compress]{natbib}
\usepackage{siunitx}
\usepackage{caption}
\usepackage[export]{adjustbox}

\def\BibTeX{{\rm B\kern-.05em{\sc i\kern-.025em b}\kern-.08em
    T\kern-.1667em\lower.7ex\hbox{E}\kern-.125emX}}
\begin{document}

\title{An RF-Domain Leakage Cancellation Scheme for FMCW Radars\\
}

\author{\IEEEauthorblockN{Yikuan Chen, Ali M. Niknejad}
\IEEEauthorblockA{
Berkeley Wireless Research Center (BWRC), University of California, Berkeley, CA
}}

\maketitle
\thispagestyle{plain}
\pagestyle{plain}

\begin{abstract}
%In Frequency-Modulated Continuous-Wave (FMCW) radars, leakage power from the transmitter (TX) to the receiver (RX) can significantly degrade the Signal-to-Noise and Distortion Ratio (SNDR), often saturating the RX front-end and limiting its dynamic range. Traditional spillover mitigation techniques, such as high-pass filtering in the base band, are ineffective for distinguishing ultra-short range targets, and do not address the problem of front-end saturation. 
This paper proposes a novel solution to address the leakage from the transmitter (TX) to the receiver (RX) in frequency-modulated continuous-wave (FMCW) radars. The proposed scheme replicates the leakage using an in-phase and quadrature mixer (IQ-mixer) and performs leakage cancellation in the radio-frequency (RF) domain. This approach utilizes a Wilkinson power combiner after the RX antenna to subtract the replicated leakage signal from the received signal, ensuring that only the true target signal reaches the low-noise amplifier (LNA). This scheme enhances the dynamic range and the receiver’s ability to discern proximate targets from previously indistinguishable low beat-frequency clutter. In addition, the proposed technique incorporates a second IQ-mixer based complex modulator in the transmitter to tune the leakage beat frequency. This allows for accurate estimation of the leakage amplitude and phase without additional hardware. Simulation results show more than 20 dB of leakage cancellation.
\end{abstract}


\section{Introduction}
\begin{figure}
\centering
\includegraphics[width=90mm]{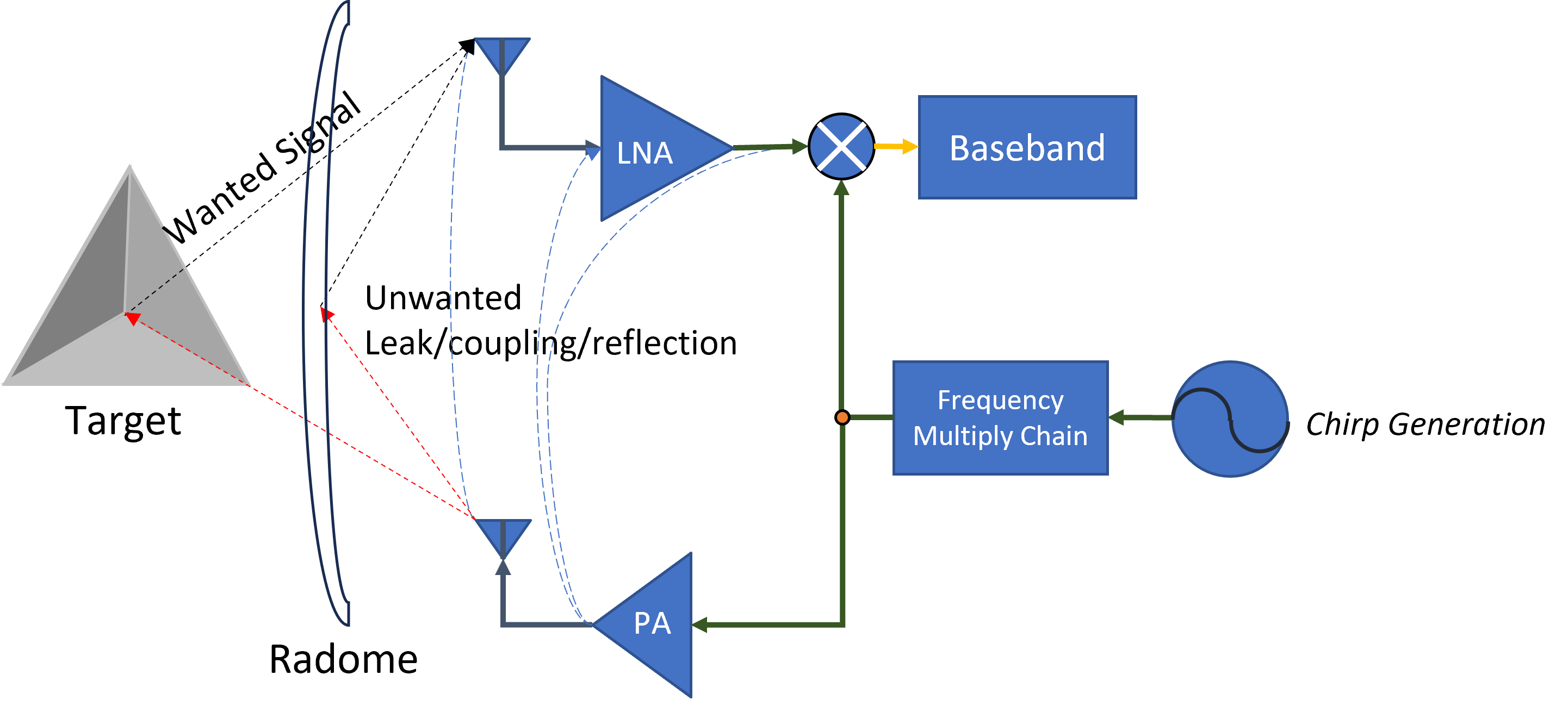}
\caption{Simplified FMCW radar architecture and self-interference paths.}
\label{fig:leakage_source}
\end{figure}

FMCW has seen significant advancements in recent years and it is playing an increasingly vital role in a wide range of applications, such as autonomous driving, security surveillance, indoor human detection, gesture recognition, biomedical monitoring systems and more. The trend of recent development of millimeter-Wave (mm-wave) FMCW radar has been toward pursuing higher dynamic range, finer range and velocity resolution, lower power consumption and interference-resilient operation. However, FMCW radar systems often suffer from TX to RX leakage, as shown in Fig. \ref{fig:leakage_source}, leading to desensitization and limited dynamic range. This is particularly problematic in short-range applications, such as gesture recognition, where the leakage results in a beat frequency that obscures nearby targets.

In long range applications such as automotive, the frequency separation of the target and the leakage typically allows for leakage mitigated in the baseband domain using high-pass filters. However, such an approach does not address the saturation problem in the RF stages. While most automotive radars operate in the 77 GHz band where high-linearity receivers are achievable with reasonable current consumption, complementary metal-oxide-semiconductor (CMOS) LNAs in higher frequency bands such as D-band struggle to maintain high linearity without excessive current consumption \cite{cmos_performance_100g}. Additionally, this approach increases the complexity of the intermediate frequency (IF)/baseband circuit and it also does not suit short-range applications where visibility in the first few Fast-Fourier Transform (FFT) bins is desired. %Additionally, the high power of the leakage imposes stringent linearity requirements on the receiver. 

\begin{figure}
\subfloat[]{%
\centering
\includegraphics[width=45mm]{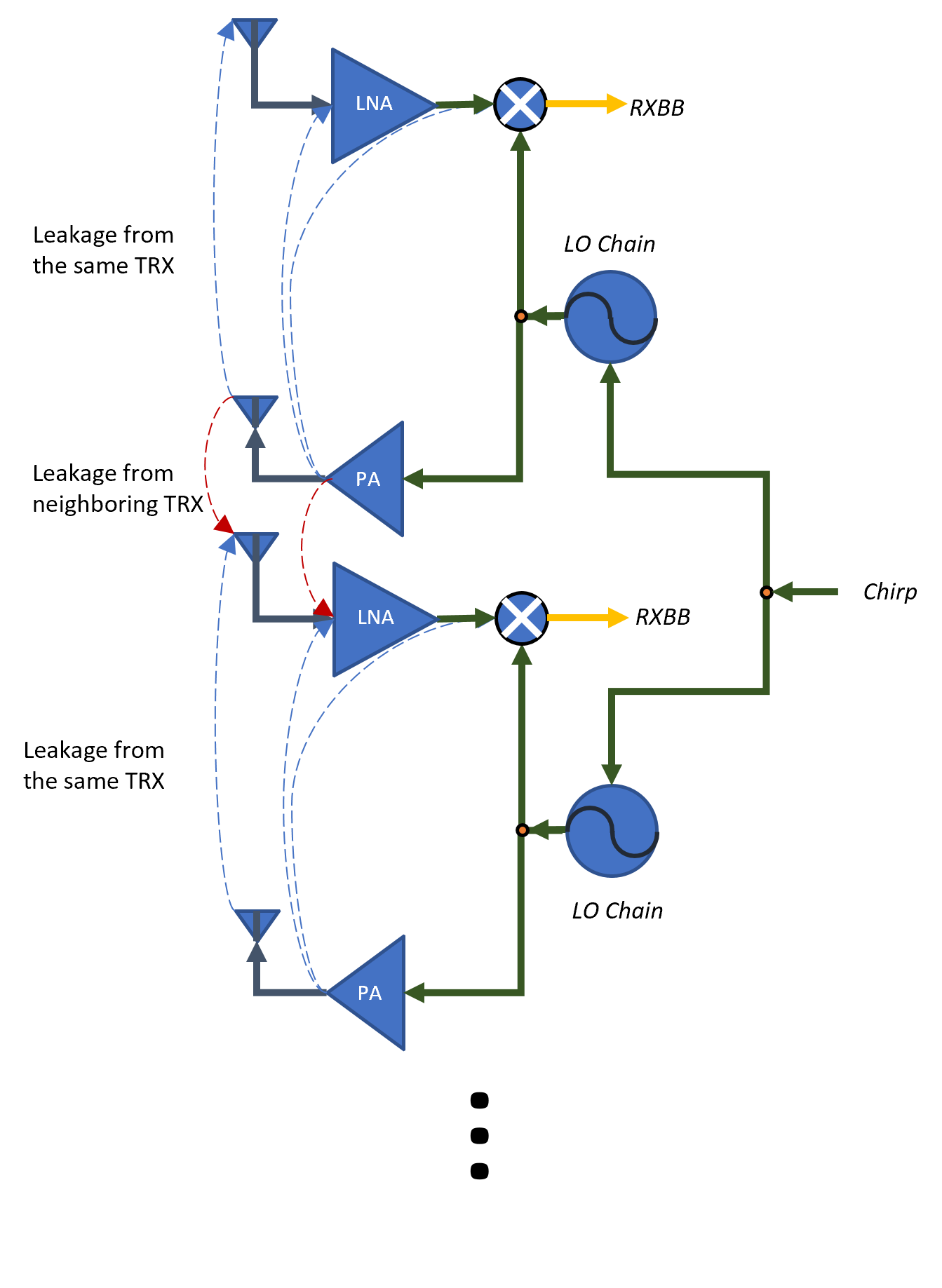}
}
\subfloat[]{%
\centering
\includegraphics[width=40mm]{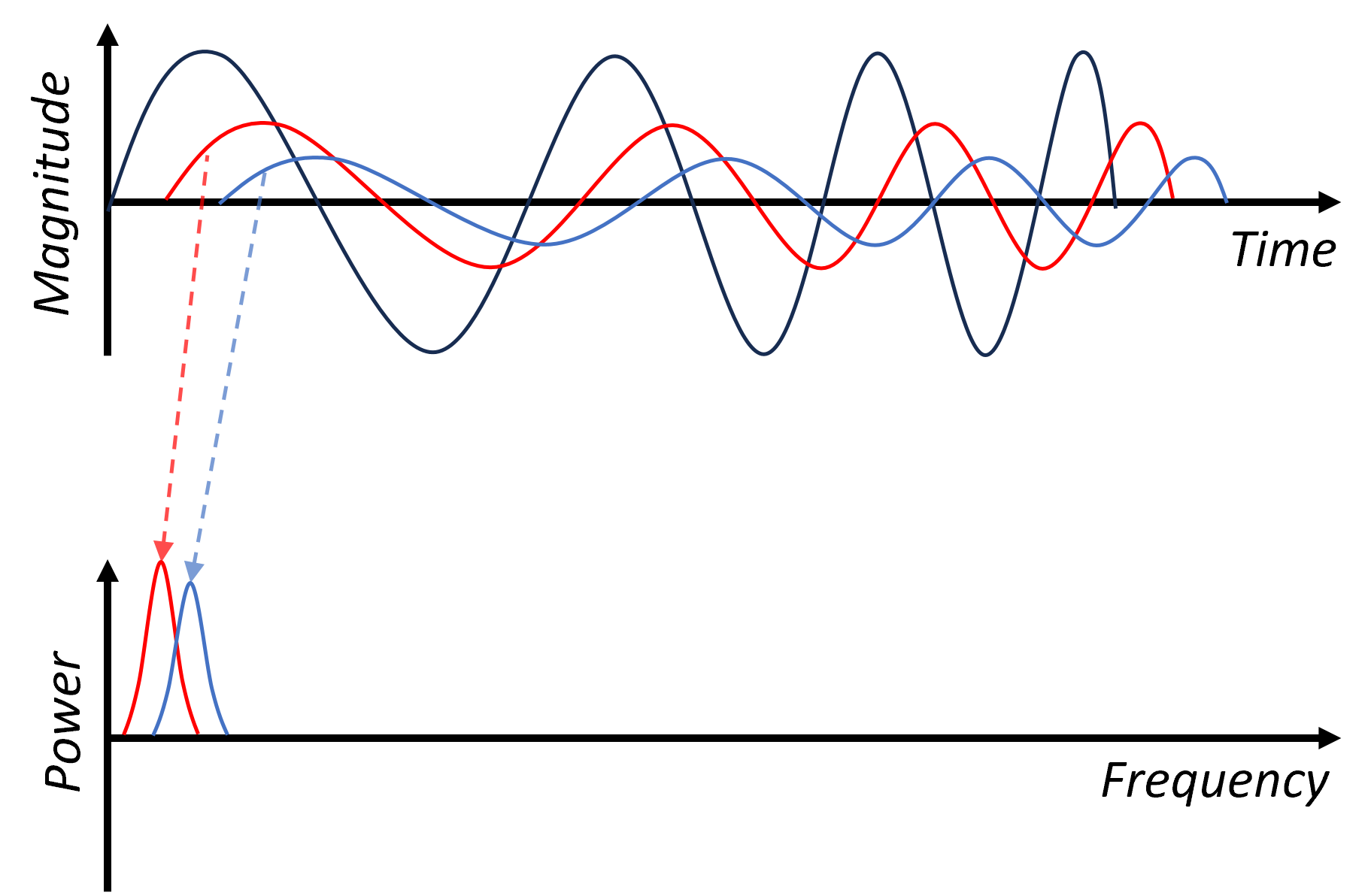}
}
\caption{(a) Self-interference from different neighboring TX elements in an array. (b) Resulting leakage signals and their beat frequency spectrum.}
\label{fig:Array}
\end{figure}

To tackle this challenge, several techniques to cancel the spillover in the RF front-end have been demonstrated. \cite{Chen_TurnstileAnt_FMCW} uses circularly polarized antenna to suppress the directly-coupled leakage. However, this approach cannot suppress short-range reflection caused by unwanted obstacles. Another common technique is to create an artificial target with matched delay to the leakage using a variable reflector \cite{Kalantar_SingleAnt_FMCW} or a vector modulator \cite{Kucharski_vecmod}. Such approaches do not address multi-path self-interference, which contains multiple beat frequency components as shown in Fig. \ref{fig:Array}.

This work introduces a novel approach to perform self-interference cancellation within the RF domain. Compared to the state-of-the-art, this technique demonstrates a significant improvement in leakage cancellation without the limitations of prior works. The proposed leakage cancellation scheme can achieve more than 20 dB of reduction prior to the RX front-end, while mitigating the multi-path self-interference. This RF leakage cancellation scheme will ease the requirement on the RX front-end linearity and simplify the radar baseband signal post-processing.

\section{Theory of FMCW Operation}
\subsection{Range Detection with FMCW Radars}
FMCW radar operates on the principle of transmitting a continuous wave signal that varies in frequency linearly over time. This variation creates a frequency sweep over certain bandwidth from \( f_{\text{start}} \) to \( f_{\text{stop}} \) over a period \( T_{chirp} \). The transmitted signal can be expressed as:
\begin{equation}\label{eq_fmcw}
    s_T(t) = A \cos \left[ 2\pi \left( f_{\text{start}}  + \frac{f_{\text{stop}} - f_{\text{start}}}{T_{chirp}} t \right)t + \phi_0\right] 
\end{equation}

Here, \( A \) is the amplitude of the transmitted signal and \(\phi_0\) is its initial phase. When the transmitted signal encounters a target, it is reflected back to the radar receiver, where it mixes with the transmitted signal, resulting in a baseband signal with beat frequency \( f_{\text{beat}} \) between the transmitted and received:

\begin{equation}\label{eq_fbeat}
 f_{\text{beat}}(t) = 2R(t)\frac{B}{c T_{chirp}} 
\end{equation}

where \(B=f_{\text{stop}} - f_{\text{start}}\), \( c \) is the speed of light and \( R(t) \) is the range of the target, equal to the half-round-trip distance traveled by the signal. By analyzing \( f_{\text{beat}}(t) \) with algorithms such as Fourier Transform (FT), of the target can be determined. However, in a real radar system, the isolation between the TX and RX is often limited, leading to a leakage signal that often has high power and low beat frequency due to the short distance between the TX and the RX.

To cancel the leakage, the delay and the amplitude of the leakage must be estimated, and a matched copy of the leakage needs to be generated and subtracted from the received signals. However, since the beat frequency of a signal in the FMCW system is proportional to the delay, the delay can be artificially synthesized by creating frequency offsets on the chirp.
% In real implementations of FMCW radar, the range resolution is limited by the bandwidth of the transmitted signal. The range resolution \( \Delta R \) is related to the bandwidth \( B = f_{stop} - f_{start} \) of the chirped signal by the equation:

% \[ \Delta R = \frac{c}{2B} \]

% This relationship shows that to achieve finer range resolution, a wider bandwidth is required. 

% To relate the range resolution to the frequency resolution, 
% Consider that the beat signal of a chirp duration is sampled at frequency \(f_s = 1/T_s\), and an \(N\)-point FFT is applied to to signal. The resulting FFT frequency bin width \(f_{bin}\) can be expressed as:

% \begin{equation}\label{eq_fbin1}
% f_{bin} = \frac{f_s}{N}
% \end{equation}

% where 

% \begin{equation}\label{eq_fbin2}
% N = \frac{T_{chirp}}{T_{s}} = f_sT_{chirp}
% \end{equation}

% and therefore

% \begin{equation}\label{eq_fbin3}
% f_{bin} = \frac{1}{T_{chirp}}
% \end{equation}

% From Eq. \ref{eq_fbeat}, we can see the linear relation ship between the beat frequency and the distance of the target. Therefore, the minimum resolvable distance, or the range resolution, of an FMCW radar can be expressed as:

% \begin{equation}\label{range_res}
% R_{bin} = \frac{f_{bin}cT_{chirp}}{2B} = \frac{c}{2B}
% \end{equation}

% Notice that the range resolution is independent of sampling frequency or chirp duration.

\subsection{Impact of Spectral Leakage}

The beat frequency of the leakage in an FMCW radar is often a fraction of the FFT bin width. Discrete Fourier Transform (DFT) theory states that if a signal or a sub-component of a signal does not have an exact integer multiple of periods within the sampled duration, spectral leakage will occur, resulting in signal energy spreading into adjacent frequency bins, distorting the true spectrum. In the frequency domain, a sampled signal with non-integer multiple period equates to the frequency not being an exact integer multiple of the DFT frequency bin size.

\begin{figure}
\subfloat[]{%
\centering
\includegraphics[width=90mm]{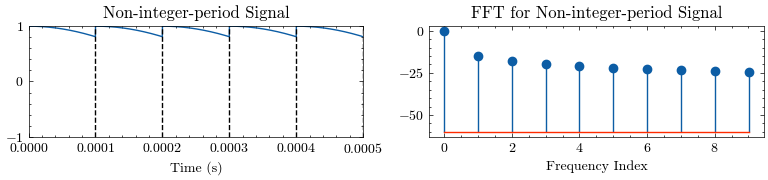}
\label{fig_noncoherent}

}
\\
\subfloat[]{%
\centering
\includegraphics[width=90mm]{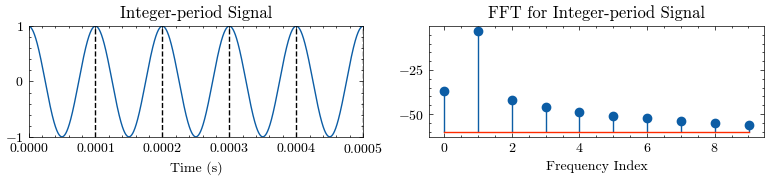}
\label{fig_coherent}
}
\caption{(a) A non-coherently-sampled signal and its FFT spectrum with spread energy. (b) A coherently-sampled signal and its FFT with energy concentrated in one bin. }
\end{figure}

Fig. \ref{fig_noncoherent} shows a beat signal with its period much longer than the chirp period. The spectral leakage imposes two challenges to the system in this case. The first challenge is the accurate amplitude estimation. Despite that the amplitude of the leakage is high in the RF domain, the de-chirped baseband signal appears as a small-amplitude signal with a large DC-offset, as the energy is spread into different frequency bins, with most energy in the DC bin. This makes the amplitude estimation of the leakage complicated as the DC-offset can also be introduced by other circuit non-idealities in the radar system. The second challenge is the detection of the true targets, as the spectral leakage in higher frequency bins can smear out the true targets because the spectral leakage can have larger magnitude than the peak of the true targets. Typically, to address the second challenge, a windowing function is applied to reduce discontinuities at frame edges and thus suppress such spectral leakages. However, this can widen the main spectral lobe which degrades the range resolution, and it does not address the first challenge. Coherent sampling, which ensures an integer number of signal cycles fit within the sampling window, eliminates the need for windowing and guarantees the full amplitude of the leakage beat signal is presented after de-chirping, as shown in Fig. \ref{fig_coherent}. When we coherently sample the leakage beat tone, it's amplitude and phase is correctly reflected from the FFT readings from a single bin. This work leverages coherent sampling of the leakage beat signal to simplify the parameter estimation of the leakage, including the frequency, amplitude, and phase.

\section{Proposed System Architecture}
\subsection{Leakage Cancellation Architecture Overview} \label{cancellation_scheme}
\begin{figure}
\centering
\includegraphics[width=90mm]{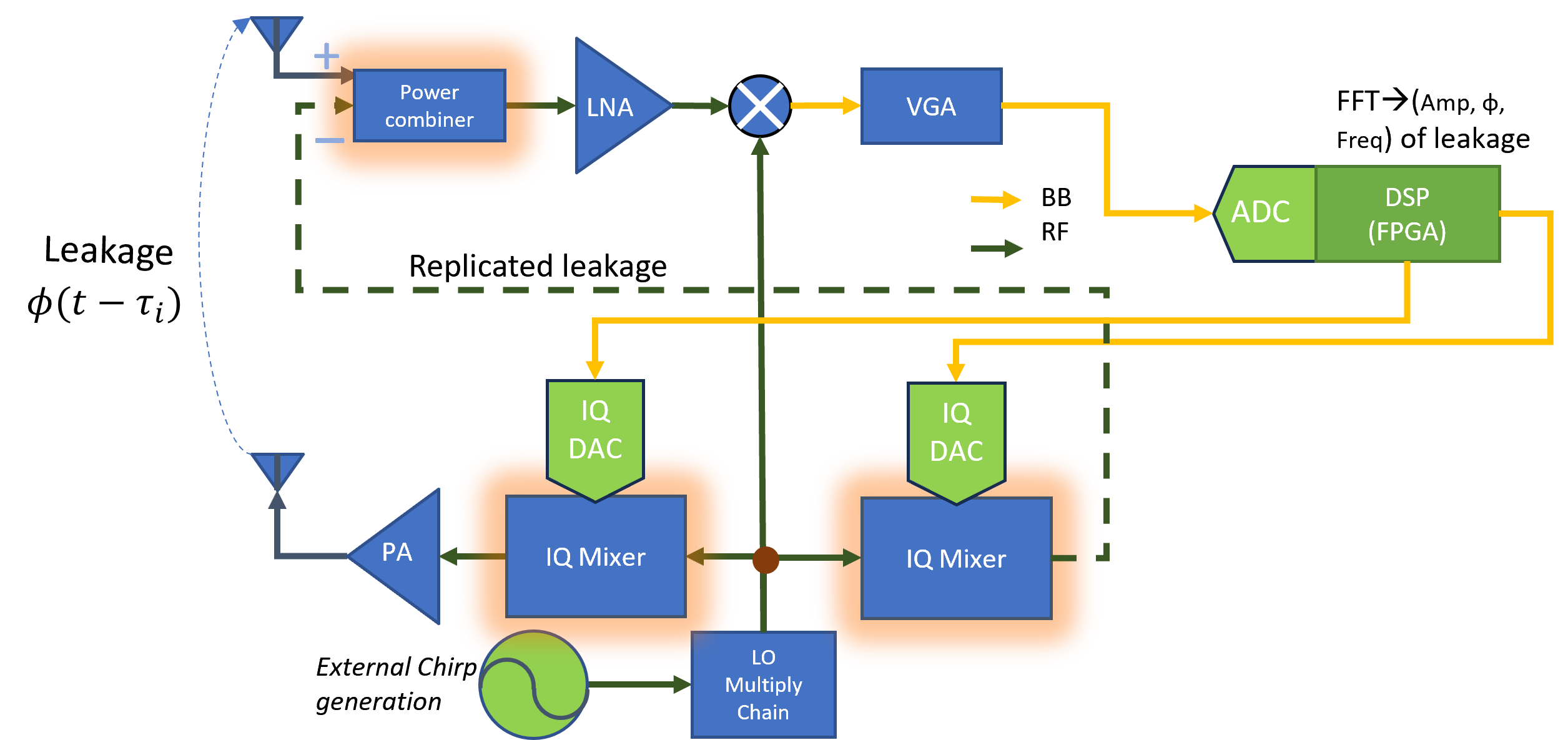}
\caption{System block diagram of the proposed leakage cancellation scheme.}
\label{fig:sys_block_diagram}
\end{figure}

% We aim to mitigate the leakage in the RF domain to ensure that short-range targets can be distinguished, and the LNA itself will also not be saturated by the leakage while operating in the highest sensitivity gain setting. 
This section introduces the proposed architecture that mitigates leakage in the RF domain to ensure short-range target distinction without saturating the LNA. The system block diagram is shown in Fig. \ref{fig:sys_block_diagram}. The three components that distinguish this work are:
\begin{enumerate}
    \item \textbf{TX IQ-mixer}: Adjusts the frequency offset of the transmitted chirp relative to the local chirp at RX de-chirping mixer to ensure coherent sampling of the leakage beat signal.
    \item \textbf{Replica IQ-mixer}: Produces a polarity-inverted version of the leakage signal.
    \item \textbf{Wilkinson Power Combiner}: Combines the received signal (leakage + desired signal) with the replicated leakage signal before the LNA stage.
\end{enumerate}

These components form a leakage-control-and-cancellation loop. The TX IQ-mixer ensures precise frequency offset adjustment. The Replica IQ-mixer generates an inverted leakage signal. The power combiner subtracts the leakage component, allowing only the desired signal to pass to the LNA. This architecture significantly enhances the radar system’s ability to discern short-range targets and reduce the impact of leakage on the receiver’s performance.
% two IQ-mixers and a power combiner. The IQ-mixer on the left adjusts the frequency offset of the transmitted chirp relative to the local reference, or the one that is sent to the RX de-chirping mixer. The IQ-mixer on the right produces the polarity-inverted version of the leakage signal, and the resulting signal is fed into one of the input port of the power combiner before the LNA stage. Together, these three components forms a leakage-control-and-cancellation loop. In the following subsections, their purpose will be explained in details.

csection{The Replica IQ-mixer}
\label{IQ_mixer_delay_line}
The role of the Replica IQ-mixer is to offset the local chirp by IQ-modulating it with a complex continuous wave (CW) tone of specific frequency, amplitude, and phase, leveraging the property of FMCW that delaying the chirp in time is equivalent to changing its offset frequency relative to the local chirp. The baseband of the Replica IQ-mixer is driven by an IQ digital-to-analog converter (DAC). This allows accurate replication of the leakage signal. Traditional vector modulator-based circuits cannot produce artificial delays for the chirp, making them unsuitable for cancelling FMCW leakage.

Additionally, leakage may consist of multiple signals taking different paths between TX and RX. The Replica IQ-mixer can synthesize the sum of multiple leakage tones to mitigate multi-path leakage.

\subsection{Leakage Cancellation-First Receiver}
To maximize the suppression of leakage power, the leakage cancellation stage is placed at the very first stage of the receiver. A Wilkinson power combiner is used to combine the received signal with the inverted leakage signal replica and provide high isolation between the Replica IQ-mixer and the RX antenna. This combination results in the subtraction of the leakage component, allowing only the desired signal to pass through. The residual signal, now with reduced leakage, is then fed into the LNA. % Due to layout constraints, only the I-channel output of the de-chirping IQ-mixer is amplified and sent off-chip through the baseband input/output (IO) pads, while the Q-channel output is grounded.

\subsection{The TX IQ-mixer}
An IQ-mixer is placed in the TX chain, similar to communication architectures. Here, this mixer also serves as an tunable artificial delay line to control the resulting baseband frequency of the dominating leakage. By tuning the baseband frequency of the TX IQ-mixer, the baseband beat frequency of the leakage can be aligned with a specific Fast Fourier Transform (FFT) bin, avoiding spectral leakage of the dominant leakage tone and maintaining high range resolution. When the spectral leakage is suppressed using this frequency alignment technique, the leakage tone parameters can be read off from the FFT results of that bin. 

By introducing the TX IQ-mixer, the system gains fine-grained control over the chirp’s beat frequency. This flexibility is crucial for managing self-interference without sacrificing range resolution due to windowing.  Furthermore, with a tunable frequency offset, the de-chirped signal can be shifted away from high-flicker noise baseband, which further improves the resulting Signal-to-Noise Ratio (SNR).

\subsection{Leakage Cancellation in an Array}
The proposed IQ-mixer approach allows each receiving element to independently adjust the parameters for different self-interference components with varying delays and magnitudes. By generating the ensemble chirp replica matching each component, the Replica IQ-mixer can simultaneously cancel self-interference at different beat frequencies originating from different TX elements in an array. In addition, as nearby array elements activate or deactivate, the proposed approach dynamically adjusts the number of tones in the ensemble baseband signal driving the Replica IQ-mixer and maintain optimal performance. 

\section{System Simulation Results}
\subsection{System Parameters}
The leakage cancellation system is modeled in MATLAB. The following link budget parameters were used: The output power from the transmitter is 10 dBm, while the coupling between TX and RX is assumed to be -30 dB, and the noise figure of the receiver is 15 dB. The chirp has a bandwidth from 135 GHz to 145 GHz and has a period of 100$\mu$s. The virtual target has -40 dBsm radar cross section (RCS), equivalent to the worst case RCS of an adult's hand \cite{hand_RCS}, is located at 10 cm away from the radar in the first case and 20 cm away in the second case. The received signal is sampled at 100 MSps with an ideal 16-bit resolution ADC, and decimated by a factor of 10. The simulation adopted 16-bit fixed point resolution for data processing. 

The frequency of the TX IQ-mixer baseband is tunable at a step of 152.6 Hz, which corresponds to a delay step of 1.53 ps. This is the finest frequency step that is achievable using a direct-digital-synthesis-based sine wave generation with 16384 samples stored on the memory for frequency generation and DAC operating at 10 MSps. This control resolution will ultimately limit the accuracy to match the cancellation signal with the received leakage signal, and hence the limit the amount of achievable cancellation. 

\subsection{Operation Steps}
The following steps are performed for the leakage estimation and cancellation.
\begin{enumerate}
    \item \textbf{TX IQ-mixer Tuning}: Adjusts the frequency offset of the transmitted chirp relative to the local reference or the RX de-chirping mixer by sweeping the frequency of the DAC driving the TX-mixer and finding the frequency corresponded to minimum spectral leakage of the first FFT after the DC bin. 
    \item \textbf{Leakage Estimation}: After the previous step, the frequency of the leakage will be equal to the resolution bandwidth (RBW) of the FFT. The amplitude and phase can be directly read from the FFT result.
    \item \textbf{Fine-tuning the Phase}: Produces a polarity-inverted version of the leakage signal on the DAC to drive the Replica IQ-mixer with the frequency, phase and amplitude obtained in the previous step. Perform a phase fine-tuning by sweeping the phase and small steps around the initial value, until the minimum leakage amplitude is achieved on the FFT spectrum.
    \item \textbf{Leakage Cancellation}: Drive the TX IQ-mixer and Replica IQ-mixer with the optimum parameters obtained in above steps. The resulting signal now contains a much lower leakage amplitude and the system's dynamic range is improved. 
\end{enumerate}

\subsection{Results}
\begin{figure}
\subfloat[]{%
\centering
\includegraphics[width=80mm]{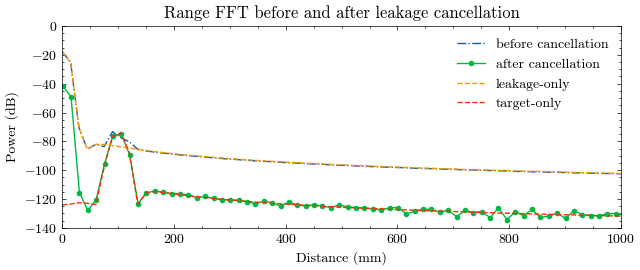}
\label{fig_leakage_cancel_10}
}

\subfloat[]{%
\centering
\includegraphics[width=80mm]{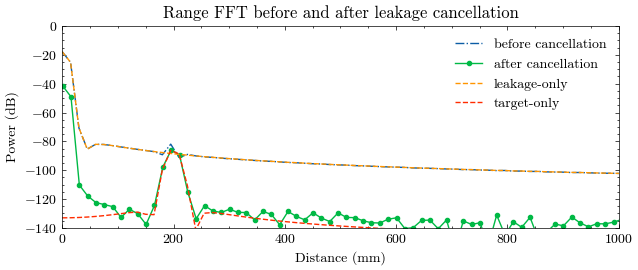}
\label{fig_leakage_cancel_20}
}
\label{fig:leakage_cancel}
\caption{(a) Simulated range FFT spectrum with and without leakage cancellation with target 10 cm away from the radar. (b) Simulated range FFT spectrum with and without leakage cancellation with target 20 cm away from the radar.}
\end{figure}

A comparison between the spectrum of the signal before cancellation with the signal after cancellation is shown Fig. \ref{fig_leakage_cancel_10}. The signal before cancellation consists of the leakage with unknown phase leakage and beat frequency as well as the target component. The signal after cancellation contains the suppressed leakage and the target component. The power of the leakage has been suppressed by 20 dB at the center of the FFT bin, and the spectral leakage from the leakage is also greatly reduced due to proper alignment of the leakage beat frequency with the FFT frequency bin. In Fig. \ref{fig_leakage_cancel_20}, where the target reflected power is much weaker due to increased distance, the target was not distinguishable before leakage cancellation, but is distinguishable after the leakage cancellation.

% \subsection{Multi-path Leakage and Leakage in an Array}

% \begin{figure}
% \centering
% \includegraphics[width=80mm]{figures/Simulations/leakage_cancel.png}
% \caption{Range FFT spectrum with and without leakage cancellation.}
% \label{fig:leakage_cancel}
% \end{figure}

\section{Concerns of the Proposed Approach}
Implementing the proposed leakage cancellation technique involves certain trade-offs and considerations: (1) Noise figure trade-off: The use of a power combiner for leakage cancellation introduces a 3dB power loss for the desired signal when the input ports are driven incoherently. This trade-off affects the overall noise figure of the receiver. (2) Loop Stability: Ensuring the stability of the cancellation loop requires thorough characterization of the leakage under varying conditions, such as temperature fluctuations and changes in group delay caused by different gain settings in the receiver. 

In short-range applications, the dynamic range of the FMCW radar is dominated by the leakage rather than the thermal noise or phase noise due to the proximity of the targets to the radar and the range correlation effect \cite{melzer}, so suppressing the leakage by accepting a the noise figure degradation can be acceptable. However, to mitigate the impact to the noise figure, a different type of directional coupler could be used, such as a 10-dB coupler. This will introduce much lower insertion loss, but also injects lower power into the receiver from the Replica IQ-mixer. Additional gain stages might be necessary after the Replica IQ-mixer to compensate for the drop in the power level. 

To enhance the stability of the loop, temperature compensation circuits (such as dynamic bias current control) or digital compensation based on pre-calibrated look up table can be added to maintain consistent leakage cancellation performance across varying temperature. Such approach requires extensive characterization under various conditions to develop robust models that can predict and adjust for changes in group delay and other parameters.

\section{Conclusion}
The proposed IQ-mixer based RF leakage cancellation scheme for FMCW radars effectively mitigates TX to RX leakage, enhancing the dynamic range and improving the radar’s ability to discern short-range targets. By leveraging precise control over the transmitted chirp’s frequency offset, the radar system can accurately assess the leakage parameters, replicate the signal in RF domain and achieve more than 20 dB of leakage cancellation. This innovative solution addresses the limitations of traditional spillover mitigation techniques and simplifies radar baseband signal processing. 
%Future work will focus on system integration, and extensive testing to validate the performance of the proposed architecture.

\section*{Acknowledgment}
% This work was supported in part by the Center for Ubiquitous Connectivity (CUbiC), sponsored by Semiconductor Research Corporation (SRC) and Defense Advanced Research Projects Agency (DARPA) under the JUMP 2.0 program. This work was supported in part by Semiconductor Research Corporation (SRC). The authors thank the TSMC University Shuttle program for chip fabrication, and Cadence for EMX.
This material is base upon work supported in part by the Center for Ubiquitous Connectivity (CUbiC), sponsored by Semiconductor Research Corporation (SRC) and Defense Advanced Research Projects Agency (DARPA) under the JUMP 2.0 program.

\small{
\bibliographystyle{IEEEtran}

\bibliography{IEEEexample}

\begin{thebibliography}{1}
\providecommand{\url}[1]{#1}
\csname url@samestyle\endcsname
\providecommand{\newblock}{\relax}
\providecommand{\bibinfo}[2]{#2}
\providecommand{\BIBentrySTDinterwordspacing}{\spaceskip=0pt\relax}
\providecommand{\BIBentryALTinterwordstretchfactor}{4}
\providecommand{\BIBentryALTinterwordspacing}{\spaceskip=\fontdimen2\font plus
\BIBentryALTinterwordstretchfactor\fontdimen3\font minus \fontdimen4\font\relax}
\providecommand{\BIBforeignlanguage}[2]{{%
\expandafter\ifx\csname l@#1\endcsname\relax
\typeout{** WARNING: IEEEtran.bst: No hyphenation pattern has been}%
\typeout{** loaded for the language `#1'. Using the pattern for}%
\typeout{** the default language instead.}%
\else
\language=\csname l@#1\endcsname
\fi
#2}}
\providecommand{\BIBdecl}{\relax}
\BIBdecl

\bibitem{cmos_performance_100g}
E.~Chou, H.~Beshary, M.~Wei, R.~Hijab, F.~Sheikh, S.~Callender, and A.~M. Niknejad, ``Comparative performance of 100–200 ghz wideband transceivers: Cmos vs compound semiconductors,'' in \emph{2023 IEEE BiCMOS and Compound Semiconductor Integrated Circuits and Technology Symposium (BCICTS)}, 2023, pp. 292--299.

\bibitem{Chen_TurnstileAnt_FMCW}
X.~Chen, M.~I.~W. Khan, X.~Yi, X.~Li, W.~Chen, J.~Zhu, Y.~Yang, K.~E. Kolodziej, N.~M. Monroe, and R.~Han, ``A 140ghz transceiver with integrated antenna, inherent-low-loss duplexing and adaptive self-interference cancellation for fmcw monostatic radar,'' vol.~65, 2022, pp. 80--82.

\bibitem{Kalantar_SingleAnt_FMCW}
M.~Kalantari, W.~Li, H.~Shirinabadi, A.~Fotowat-Ahmady, and C.~P. Yue, ``A w-band single-antenna fmcw radar transceiver with adaptive leakage cancellation,'' \emph{IEEE Journal of Solid-State Circuits}, vol.~56, no.~6, pp. 1655--1667, 2021.

\bibitem{Kucharski_vecmod}
M.~Kucharski, W.~A. Ahmad, H.~J. Ng, and D.~Kissinger, ``Monostatic and bistatic g-band bicmos radar transceivers with on-chip antennas and tunable tx-to-rx leakage cancellation,'' \emph{IEEE Journal of Solid-State Circuits}, vol.~56, no.~3, pp. 899--913, 2021.

\bibitem{hand_RCS}
P.~Hügler, M.~Geiger, and C.~Waldschmidt, ``Rcs measurements of a human hand for radar-based gesture recognition at e-band,'' in \emph{2016 German Microwave Conference (GeMiC)}, 2016, pp. 259--262.

\bibitem{melzer}
A.~Melzer, A.~Onic, F.~Starzer, and M.~Huemer, ``Short-range leakage cancelation in fmcw radar transceivers using an artificial on-chip target,'' \emph{IEEE Journal of Selected Topics in Signal Processing}, vol.~9, no.~8, pp. 1650--1660, 2015.

\end{thebibliography}
}

\end{document}